# Large Language Models vs. Search Engines: Evaluating User Preferences Across Varied Information Retrieval Scenarios

Kevin Matthe Caramancion
*Mathematics, Statistics, and Computer Science Department*
*University of Wisconsin–Stout*
Menomonie, Wisconsin, United States
caramancionk@uwstout.edu / www.kevincaramancion.com

*Abstract*—This study embarked on a comprehensive exploration of user preferences between Search Engines and Large Language Models (LLMs) in the context of various information retrieval scenarios. Conducted with a sample size of 100 internet users (N=100) from across the United States, the research delved into 20 distinct use cases ranging from factual searches, such as looking up COVID-19 guidelines, to more subjective tasks, like seeking interpretations of complex concepts in layman's terms. Participants were asked to state their preference between using a traditional search engine or an LLM for each scenario. This approach allowed for a nuanced understanding of how users perceive and utilize these two predominant digital tools in differing contexts. The use cases were carefully selected to cover a broad spectrum of typical online queries, thus ensuring a comprehensive analysis of user preferences. The findings reveal intriguing patterns in user choices, highlighting a clear tendency for participants to favor search engines for direct, fact-based queries, while LLMs were more often preferred for tasks requiring nuanced understanding and language processing. These results offer valuable insights into the current state of digital information retrieval and pave the way for future innovations in this field. This study not only sheds light on the specific contexts in which each tool is favored but also hints at the potential for developing hybrid models that leverage the strengths of both search engines and LLMs. The insights gained from this research are pivotal for developers, researchers, and policymakers in understanding the evolving landscape of digital information retrieval and user interaction with these technologies.

## I. Introduction

In the rapidly evolving landscape of information technology, the power and potential of search engines and large language models (LLMs) have captivated the attention of users worldwide. As these two dominant tools compete for primacy in information retrieval, understanding user preferences when faced with a multitude of use cases becomes paramount. This study seeks to shed light on this critical juncture in technology, offering a comprehensive analysis of user choices when presented with the options of utilizing either search engines or LLMs across various scenarios. The exploration is guided by the following research questions:

RQ1: When confronted with a range of information needs, from the latest health guidelines to understanding complex scientific concepts, do users prefer the capabilities of a search engine or the conversational interaction offered by a large language model?

RQ2: What patterns emerge in user preferences for search engines versus LLMs when the tasks involve diverse activities such as learning new skills, seeking advice, or engaging with cultural and historical content?

Search engines and large language models (LLMs) represent the twin pillars of modern information retrieval, each offering distinct advantages and limitations. Search engines, the long-established sentinels of the internet, function through indexing [1] vast quantities of data and providing users with a comprehensive list of resources in response to their queries. They serve as gateways to the internet's expanse, capable of returning a multitude of perspectives on any given topic. On the other hand, LLMs, epitomized by their conversational prowess and deep learning [2] capabilities, offer a more interactive and human-like engagement, synthesizing information and presenting it in a contextually relevant manner. This dynamic interplay between the algorithmic precision of search engines and the evolving intelligence of LLMs forms the crux of this study.

In an era marred by misinformation and disinformation [3], the authority wielded by an information source in affirming what is accurate cannot be overstated. There is power, indeed, in a source telling what's correct or not, especially in the misinformation and disinformation age. It is this very power that defines the trust and reliability placed in these tools by users. Search engines, with their ability to pull from an extensive and diverse array of content, offer a breadth of information that is unparalleled. Yet, the responsibility to sift through and evaluate the credibility of this information falls upon the user. Conversely, LLMs, with their sophisticated algorithms, promise to distill and clarify content, potentially guiding users more directly to credible information. However, they too are not immune to inaccuracies and biases [4], reflecting the challenges inherent in training AI with vast and varied datasets.

As this paper juxtaposes the capabilities of search engines against LLMs, it seeks to illuminate the nuances of each approach and how users navigate these waters. The comparative analysis conducted herein is not merely an academic exercise; it has real-world implications for how we consume and trust the information that shapes our perceptions, decisions, and understanding of the world. By examining the preferences of internet users within a controlled study, this paper aims to contribute a significant chapter to the narrative of digital information discovery.

This paper aims to contribute to the discourse on digital information retrieval by providing empirical insights into the decision-making processes of users. It also aspires to inform developers, researchers, and policymakers about the contexts in which one tool is favored over the other, potentially shaping the future development of these technologies.

The structure of this paper is meticulously designed to facilitate a clear understanding of our methodology and findings. Following this introduction, we proceed into the background necessary for understanding the nuances of the comparative study. We then explain our research methodology, enabling reproducibility and transparency. Subsequent sections present our findings, discuss their implications in depth, and explore the potential limitations of our approach. The conclusion not only synthesizes our insights but also charts a course for future research in this rapidly evolving area of inquiry.

## II. LITERARY BACKGROUND

### A. Evolution and Influence of Search Engines in Information Retrieval

Search engines are the cornerstone of modern information discovery, evolving from simple keyword-matching systems [5] to complex algorithms capable of understanding user intent [6]. They work by scanning the internet to create an index of information, which they sort and present in response to user queries. The development of these platforms has been instrumental in shaping user behavior and expectations regarding information access. The discussion will include the intricacies of Search Engine Optimization (SEO) [7], which refers to practices aimed at increasing the visibility of web content in search results, and the critical role of search engines in curating reliable information, an important defense against the spread of misinformation and disinformation.

### B. Emergence and Capabilities of Large Language Models (LLMs)

LLMs such as OpenAI's GPT-3 have introduced an interactive dimension to digital information through their capacity to understand and generate human-like text. The capability of these models to engage in dialogue and produce coherent responses is rooted in advanced machine learning techniques, including natural language processing (NLP) [8]. Their potential applications, alongside their limitations, such as biases from training data, will be examined to provide a balanced view of this technology's impact on information retrieval.

### C. The Role of Information Sources in the Misinformation and Disinformation Age

In the digital age, the credibility of an information source is paramount, given the rapid spread of misinformation and disinformation. This analysis will look into the societal implications of these phenomena, exploring how trust in search engines and LLMs shapes public perception. Both are viewed as arbiters of truth, tasked with the challenge of discerning and validating factual content for users [4][9].

### D. User Preference Dynamics: Choosing Between Search Engines and LLMs

User preferences for search engines or LLMs are influenced by various factors, including the nature of the information sought, the context of the query, and the user's familiarity with the technology [10]. Exploring these preferences involves analyzing user satisfaction in relation to the tools' speed, accuracy, and overall experience [11]. The perceived reliability and ease of use of these information sources play a significant role in their selection.

### E. Comparative Performance in Specific Use Cases

Assessing the performance of search engines and LLMs across different scenarios is crucial to understanding their respective strengths and limitations. Studies that have revealed how each of these perform a wide array of tasks such as academic research to daily problem-solving, which reveal a complex picture of their effectiveness. The focus of this paper will be on evaluating these tools in situations that demand precision, depth, and expedience.

### F. The Literary Gap: Synthesizing Search Engines and LLMs in User Preference Research

While research abounds on the individual characteristics of search engines and LLMs, a direct comparison of these technologies from the user's perspective remains sparse, particularly across a diverse array of use cases. Highlighting the importance of bridging this gap, the discussion will advocate for a holistic approach that not only compares the technical competencies of these tools but also integrates the user's experience and situational context into the analysis.

## III. METHODS

### A. Research Design

This study employed a quantitative approach, focusing on a survey conducted with a sample of 100 internet users from across the United States. The objective was to investigate the preferences between search engines and large language models (LLMs) in 20 different use cases, representing a broad spectrum of information retrieval scenarios.

### B. Participant Sampling

The survey engaged 100 participants, selected through random sampling to represent a diverse cross-section of the U.S. internet user population. This sample size was chosen to provide a snapshot of user preferences while ensuring a manageable and focused analysis. The participants varied in age, education level, technology proficiency, and frequency of technology use, offering a comprehensive overview of different demographic groups.

*C. Survey Structure and Use Case Scenarios*

Participants were presented with 20 use case scenarios in the survey, each designed to assess the preference for search engines or LLMs in specific information-seeking contexts. The scenarios covered a range of topics, including health, technology, finance, and education, to encapsulate the diverse nature of online search queries.

*D. Data Collection and Presentation*

   *1) Descriptive Statistics*

Presenting the proportion of participants who preferred search engines or LLMs for each use case.

   *2) Demographic Observation*

While not employing inferential statistics, the study will report any noticeable demographic trends in preferences, if present.

   *3) Comparative Analyses*

A comparative overview of the preferences for search engines versus LLMs across the different scenarios will be provided to identify any patterns or trends in user choices.

*E. Limitations and Scope*

   *1) Self-Reporting Bias*

The study relies on self-reported preferences, which might be subject to personal biases.

   *2) Sample Size and Scope*

With a sample size of 100 participants, the study provides insights specific to the sample group, and the results may not be indicative of broader trends among all U.S. internet users.

   *3) Geographical Focus*

The study is limited to participants from the United States, hence the findings may not be applicable to other regions.

   *4) Non-Inferential Nature*

The study is descriptive and does not engage in inferential statistical analysis, focusing instead on presenting the survey findings as observed.

*F. Data Sharing and Accessibility*

Upon completion of the study and analysis, the resulting dataset, encompassing all the survey responses and observations, will be uploaded to Kaggle, a popular platform for data science and machine learning. This decision is in line with the principles of open science and data transparency. Making the dataset publicly available on Kaggle will allow other researchers, data scientists, and interested parties to access, review, and utilize the data for further research or analysis. This step aims to contribute to the collective knowledge base and facilitate additional studies in the field of information retrieval preferences. The dataset will include detailed information from the survey while ensuring that all personal and identifiable information of the participants is removed or anonymized to maintain confidentiality and adhere to ethical standards.

## IV. RESULTS & FINDINGS

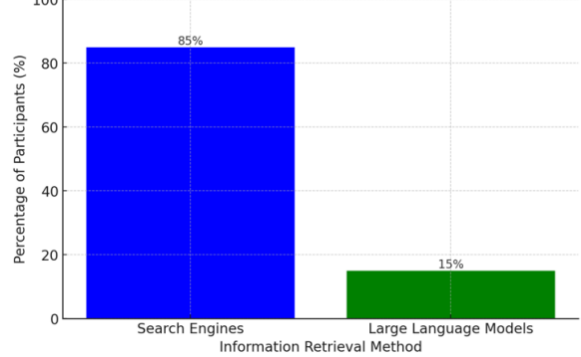

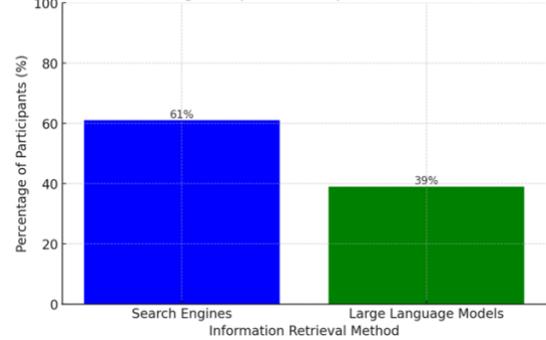

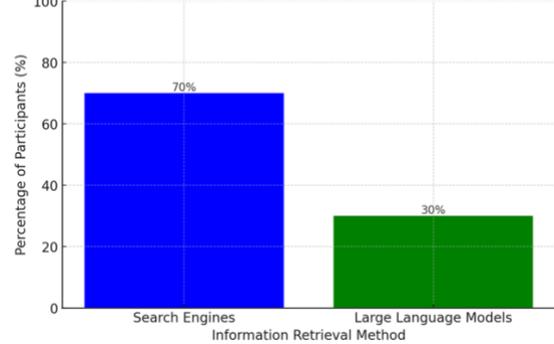

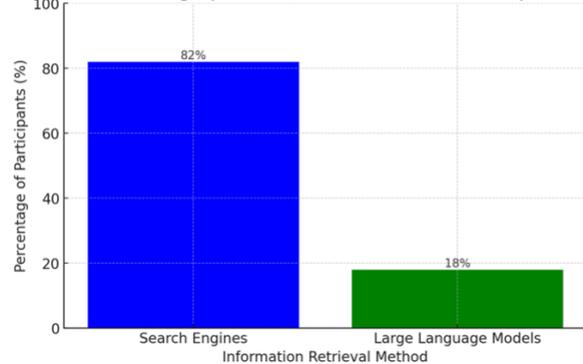

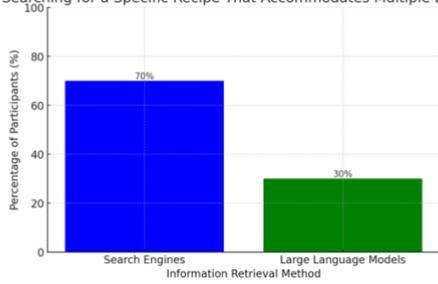

Use Case 5: Searching for a Specific Recipe That Accommodates Multiple Dietary Restrictions
- Search Engines: 70%
- Large Language Models: 30%

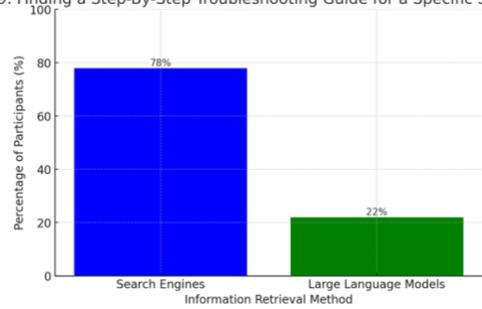

Use Case 9: Finding a Step-By-Step Troubleshooting Guide for a Specific Smartphone Issue
- Search Engines: 78%
- Large Language Models: 22%

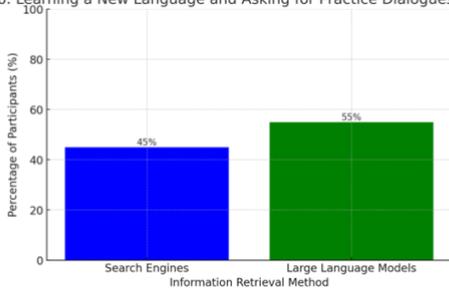

Use Case 6: Learning a New Language and Asking for Practice Dialogues and Corrections
- Search Engines: 45%
- Large Language Models: 55%

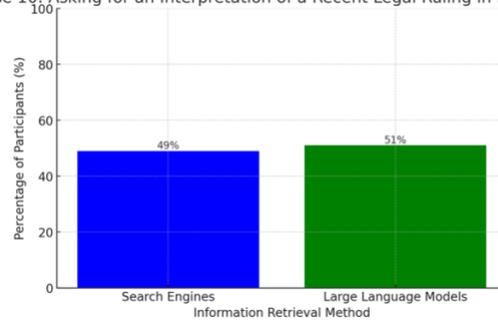

Use Case 10: Asking for an Interpretation of a Recent Legal Ruling in Layman's Terms
- Search Engines: 49%
- Large Language Models: 51%

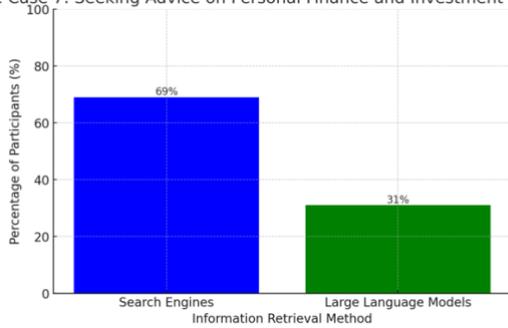

Use Case 7: Seeking Advice on Personal Finance and Investment Strategies
- Search Engines: 69%
- Large Language Models: 31%

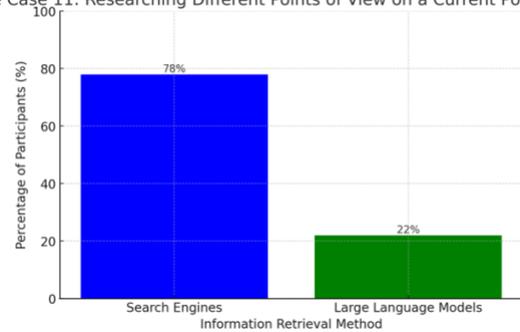

Use Case 11: Researching Different Points of View on a Current Political Topic
- Search Engines: 78%
- Large Language Models: 22%

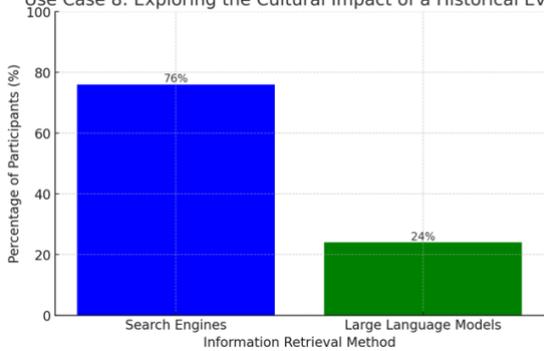

Use Case 8: Exploring the Cultural Impact of a Historical Event
- Search Engines: 76%
- Large Language Models: 24%

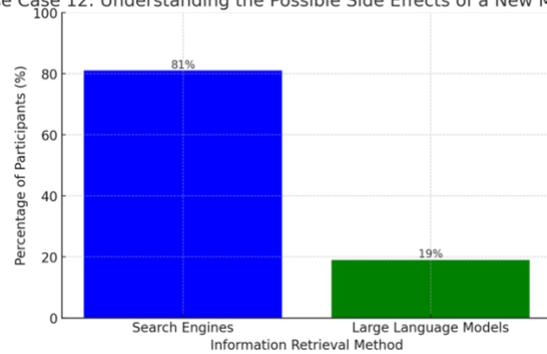

Use Case 12: Understanding the Possible Side Effects of a New Medication
- Search Engines: 81%
- Large Language Models: 19%

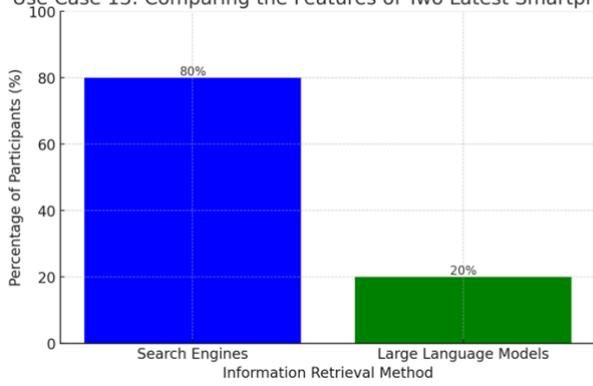
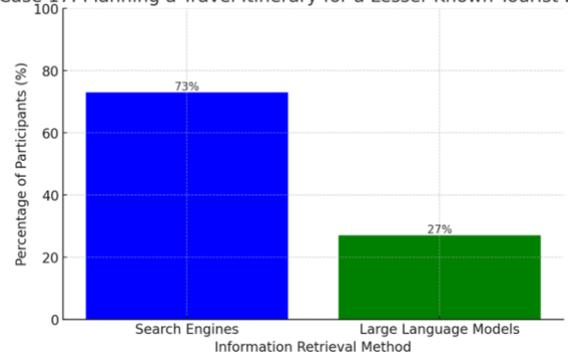
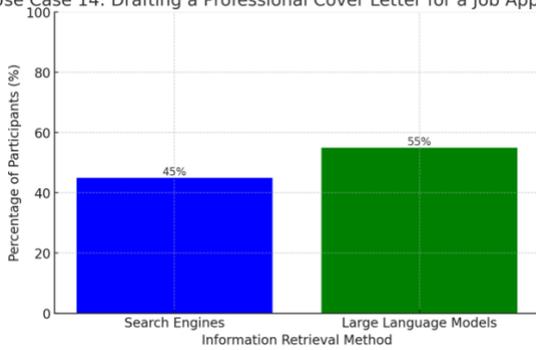
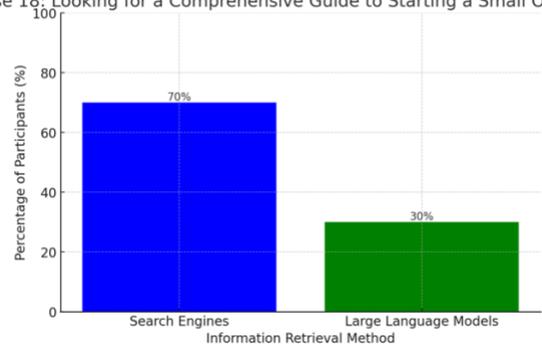
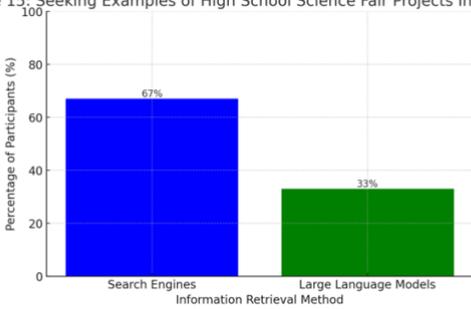
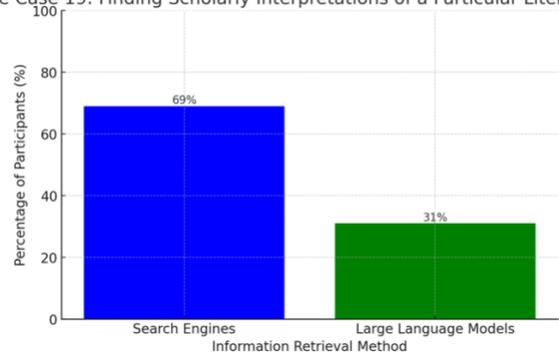
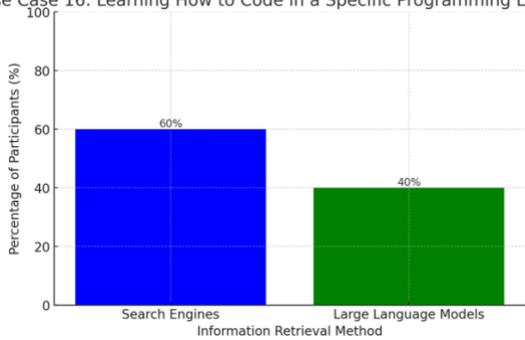
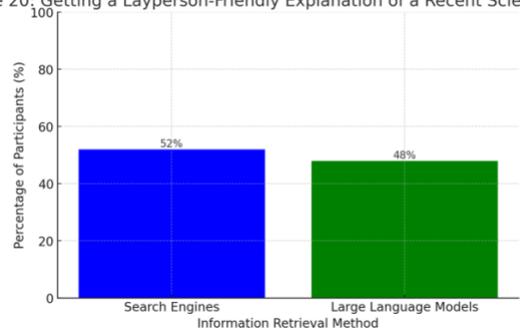

## V. ANALYSES AND DISCUSSION POINTS

*Discussion 1: Search Engines Dominate in Factual Information Retrieval*

Response to Findings: The study's results indicate a clear preference for search engines in scenarios requiring factual data retrieval, such as finding COVID-19 safety guidelines or historical weather data. Search engines were favored by a significant majority of users in these cases, with preferences often exceeding 70%. This suggests that users trust the comprehensive indexing and quick retrieval capabilities of search engines for objective, fact-based queries. The vast databases and sophisticated algorithms of search engines enable users to access a wide array of information sources, providing a breadth of data that is currently unparalleled by LLMs.

*Discussion 2: LLMs Preferred for Subjective and Language-Related Tasks*

Response to Findings: LLMs showed a stronger preference in scenarios involving language learning and seeking interpretations in layman's terms. This preference aligns with the strengths of LLMs in understanding context, generating human-like responses, and processing natural language. The close-to-even split in preferences for these use cases suggests that users recognize and value the conversational and nuanced understanding capabilities of LLMs, especially in tasks requiring a more personalized and interpretive approach.

*Discussion 3: Navigational Ease vs. Conversational Depth*

Response to Findings: The study highlights a distinct division in user preferences, with search engines being favored for their navigational ease in accessing a wide range of information and LLMs preferred for their depth in conversational interaction and nuanced understanding. This dichotomy suggests that while search engines are seen as gateways to vast information, LLMs are increasingly being recognized for their ability to simplify complex information and provide more engaging, human-like interactions.

*Discussion 4: The Growing Role of LLMs in Complex Queries*

Response to Findings: Despite the overall dominance of search engines, LLMs are closing the gap in scenarios involving complex queries, such as understanding scientific concepts or exploring cultural impacts. The nearly even split in preferences for some of these cases indicates a growing user confidence in LLMs' capabilities to handle complex, multifaceted questions that require more than just factual data retrieval.

*Discussion 5: Implications for Future Development of Information Retrieval Tools*

Response to Findings: The findings underscore the evolving landscape of digital information retrieval, where both search engines and LLMs play crucial but distinct roles. This suggests a future where the development of these tools might focus on leveraging their respective strengths—search engines for their vast, factual database access, and LLMs for their advanced natural language processing and conversational abilities. The study also highlights the potential for hybrid models that combine the factual accuracy and breadth of search engines with the contextual understanding and conversational depth of LLMs.

## VI. CONCLUSION & FUTURE WORKS

This research has provided essential insights into the current state of digital information retrieval, highlighting the distinct roles and preferences for search engines and large language models (LLMs) among internet users. As a response to the RQ1, findings indicate a clear division: users tend to prefer search engines for straightforward, fact-based queries, while LLMs are more favored for tasks requiring natural language processing and personalized responses.

The conclusion from this study is not just an academic observation but a vital guidepost for future technological development. As an explicit response to RQ2, the result of this study suggests that while search engines remain the go-to resource for quick, factual information, there is a growing appreciation and utility for the conversational and context-aware capabilities of LLMs. This trend points to a future where the integration of these technologies could provide a more holistic and efficient information retrieval experience.

Future research should explore the development of systems that combine the data-rich, comprehensive capabilities of search engines with the intuitive, conversational nature of LLMs. Such advancements could lead to more nuanced and user-friendly search experiences, catering to a wider range of information needs.

Additionally, these findings underscore the need for further investigation into user behaviors and preferences in digital information retrieval. Understanding why users choose certain tools or methods in specific contexts can inform the design of more effective and user-centric search technologies.

In conclusion, this study marks a step forward in our understanding of digital information retrieval, paving the way for future innovations. As technology continues to evolve, the focus should be on creating tools that not only enhance access to information but also enrich the overall experience of acquiring knowledge. This balance between technological advancement and user-centric design will be crucial in shaping the future of how we interact with information in the digital age.